\documentclass[runningheads]{llncs}
\usepackage{graphicx}
\usepackage{amsmath}
\usepackage{amsfonts}
\usepackage{placeins}
\DeclareMathOperator{\Tr}{Tr}
\begin{document}
\title{Evaluating glioma growth predictions as a forward ranking problem }
%
%
\author{Karin A. van Garderen\inst{1} \and
Sebastian R. van der Voort\inst{1} \and
Maarten M.J. Wijnenga\inst{1} \and
Fatih Incekara \inst{1, 2} \and
Georgios Kapsas\inst{1} \and
Renske Gahrmann \inst{1} \and
Ahmad Alafandi \inst{1} \and
Marion Smits\inst{1} \and
Stefan Klein\inst{1}}
\authorrunning{K.A. van Garderen et al.}
%
\institute{Department of Radiology and Nuclear Medicine, Erasmus MC, Rotterdam \and
Department of Neurosurgery, Erasmus MC, Rotterdam}

\maketitle
%
\begin{abstract}
    The problem of tumor growth prediction is challenging, but promising results have 
    been achieved with both model-driven and statistical methods. In this work,
    we present a framework for the evaluation of growth predictions that focuses on 
    the spatial infiltration patterns, and specifically evaluating a prediction of future growth. 
    We propose to frame the problem as a ranking problem rather than a segmentation problem.
    Using the average precision as a metric, we can evaluate the results with segmentations while
    using the full spatiotemporal prediction. Furthermore, by separating the model goodness-of-fit from future predictive performance,
    we show that in some cases, a better fit of model parameters does not guarantee a better the predictive power.
\end{abstract}
\section{Introduction}
As the diagnosis and delineation of glioma has improved with machine learning \cite{Bakas2018}, researchers look towards
the more challenging task of predicting the disease trajectory into the future \cite{Petersen2019,Elazab2020}. However, the 
problem of tumor growth is challenging in many ways, not just by the lack of publicly available data. 
The variables of clinical importance, such as the speed of infiltration and proliferation, are unknown and 
the problem of estimating them from observations is ill-posed. Furthermore, the observations we do have 
are flawed as tumor cells are known to spread beyond the visible boundary on MR imaging \cite{Silbergeld1997}, which poses 
a challenge not just in prediction but also in validation of that prediction.

Despite these challenges, biophysical growth models have shown promise in their ability to 
predict the spatial growth patterns for individual cases. They are model-driven rather than data-driven,
and strongly rooted in a mechanistic understanding of tumor growth. Delineations of the tumor on MR imaging typically
form the input for individual model fitting, with follow-up imaging providing the gold standard of evaluation. 
Though other methods of evaluation exist, such as biopsy samples \cite{Gaw2019} or PET imaging \cite{Lipkova2019}, for most 
clinical cases consecutive delineations are the best approximation for a ground truth. 

Due to the nature of the data, growth predictions are often framed as a segmentation problem. 
For example, by using an overlap metric such as the Dice Similarity Coefficient based on a sample in time 
\cite{Elazab2017,Petersen2019}. Although this metric comes natural to the ground-truth data, it 
is less representative of the underlying problem.
The main disadvantage of overlap-based metrics 
is that they treat all voxels equally, while some errors are more significant than others. 
Intuitively, we would want to assign more significance to false negative predictions at a large distance to 
the predicted tumor boundary as they are less in agreement to the model. 
This intution is represented in metrics based on the segmentation boundary, such as the symmetric surface distance 
used in Konukoglu et al. \cite{Konukoglu2010}. But even a distance metric compares only to a single point in time, and 
using a boundary metric becomes less appropriate when the ground truth contains new disconnected lesions. 

Another challenge in the evaluation of tumor growth predictions is the entanglement of model fit and prediction. All tumor 
growth models require an initial observation to fit model parameters. The segmentation on this initial observation serves as 
a measure for the goodness-of-fit and the prediction is performed from 
the time of onset, through the initial observation towards the future \cite{Angelini2007}. This work
explores the distinction between goodness-of-fit and predictive performance by strictly separating them in the evaluation.

In this work we propose the following contributions:
\begin{enumerate}
    \item A novel framing of tumor growth as a ranking problem, with the Average Precision as the performance metric
    \item An experimental design that separates the prediction performance from the goodness-of-fit in both time and space
    \item The application of this evaluation framework on a biophysical tumor growth model and a dataset of 21 patient cases, with low-grade glioma and partial tumor resection
\end{enumerate}

\section{Problem definition}
\subsection{Tumor growth as a ranking problem}
To define the problem we must first define the shape of the expected solutions. 
We are interested in predicting infiltrative growth in a spatial sense. In other words, to simplify the problem we assume that 
the speed of growth and potential mass effect are not of interest. 
We assume that a growth model could produce a segmentation of the tumor $S(t)$ at any time 
$t > 0$. It may therefore assign to every location in the brain a time $T(x)$, which is the first time $t$ when 
the tumor reaches that location. As we do not require an accurate estimation of the growth speed, we require only that the 
estimated $T(x)$ is a ranking of voxels in the brain, such that:

\begin{equation}
T(x_a) > T(x_b) \Leftrightarrow \exists t : x_a \notin S(t), x_b \in S(t).
\end{equation}

The ranking can be evaluated by a sampling of the ground-truth segmentation $S'$, by using the Average Precision (AP).
The AP is defined as the area under the Precision-Recall (PR) curve:

\begin{equation}
AP = \Sigma_t (R(t) - R(t-1))P(t),
\end{equation}

where $R(t)$ and $P(t)$ are the recall and precision at a threshold $t$ on the 
time-to-invasion ranking $T$, comparing to the segmentation $S'$. Since the precision scores are weighted by the difference in recall,
all tumor volume predictions are taken into account from the tumor onset to the time when the recall is 1, 
meaning that the ground-truth segmentation is completely encompassed by the prediction. An evaluation 
based on a single time $t$ would represent a point on the PR curve. If we take a volume-based sample, where the 
estimated tumor volume equals the observed tumor volume, this is the time $t$ where $R(t) = P(t)$.

\begin{figure*}[htb]
    \begin{center}
    \includegraphics[width=\linewidth]{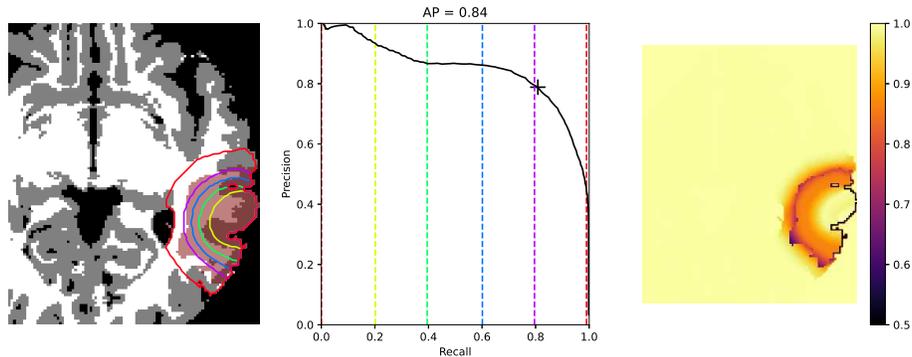}
    \caption{Left: tissue segmentation of a specific case with thresholds on the $T$ map, generated by a tumor growth model,
     indicated as segmentation boundaries. The ground-truth segmentation $S$ is indicated by a red overlay. Middle: 
    corresponding Precision-Recall curve with the same thresholds indicated. The sample with a corresponding volume is marked on the PR curve. 
    Right: quantification of agreement by $R(T(x))$ outside $S$ and $P(T(x)))$
    for voxels inside $S$.}
    \label{fig:pr_curve}
    \end{center}
\end{figure*}

Formulating the problem as a ranking and using the AP has a number of qualitative advantages. First, the ranking $T$ has 
a direct local connection to the speed of the tumor boundary. If the ranking is smooth, the gradient of the $T$ represents the local
movement of the visible tumor boundary. And yet, as a second advantage, the AP metric does not impose any assumption of smoothness on $T$. 
It automatically assigns a larger weight to certain parts of the prediction, depending on the assigned ranking $T$, 
regardless of any assumptions on the significance of distance in space or time. We might quantify the agreement between $T$ and $S$ locally 
by using the rank of the voxel $T(x)$ as a threshold on the PR curve. A local prediction $T(x)$ is in 
agreement with $S$ if it is part of the reference segmentation ($x \in S$) and can be included 
with high precision $P(T(x))$, or else if it falls outside $S$ but can be excluded with high recall $R(T(x))$.
Figure \ref{fig:pr_curve} illustrates the computation of the AP metric and this local measure of disagreement.

\subsection{Experimental design}
\begin{figure*}[htb]
    \includegraphics[width=\linewidth]{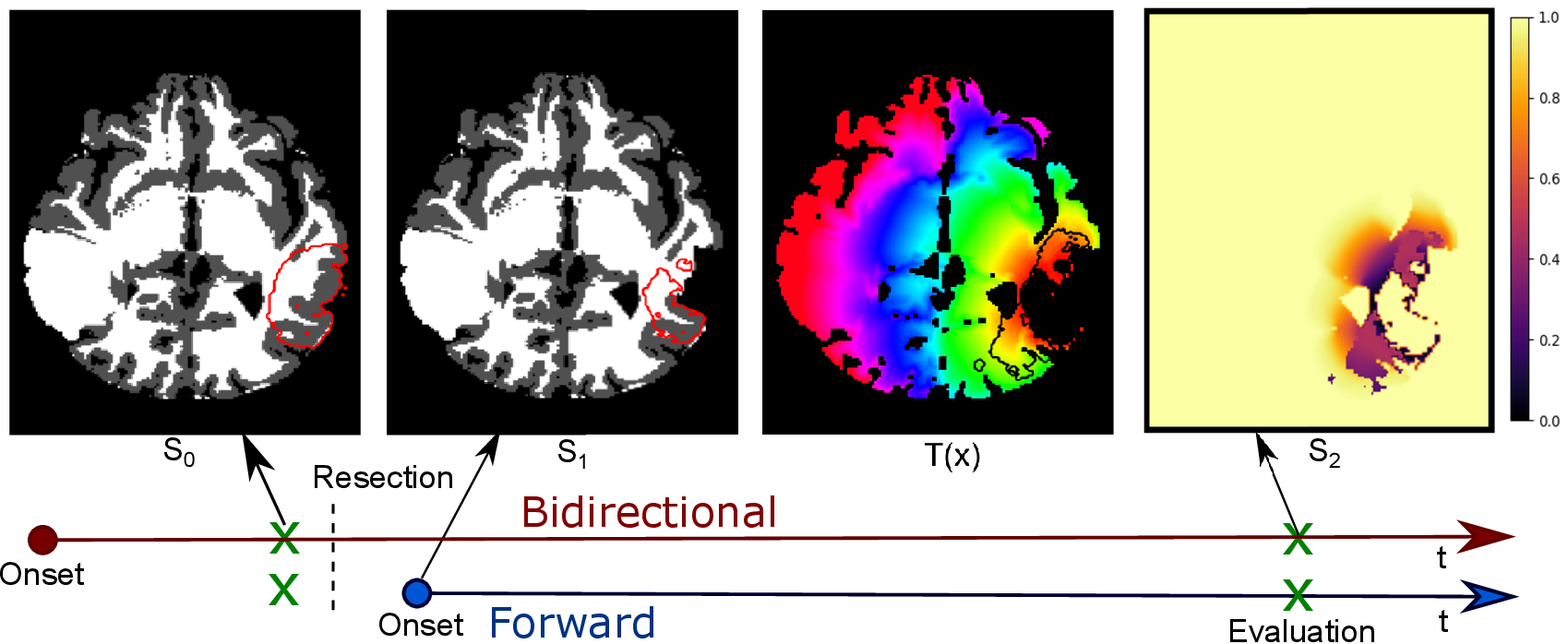}
    \caption{Overview of two temporal evaluation schemes. Bidirectional: a growth model is fitted to the initial tumor
    and simulated from a seed point to generate a voxel ranking $T$. Forward: parameters are fitted to the initial tumor
    and then the model is initialized with a segmentation $S_1$ obtained after resection to generate the voxel ranking $T$. Images from 
    left to right: example of tissue segmentation with $S_0$ outlined, tissue segmentation with resection cavity removed and $S_1$ outlined,
    example of final ranking $T$ used for the evaluation 
    with resection cavity and $S_1$ removed, quantification of agreement between $T(x)$ and $S_2$   }
    \label{fig:timeline}
\end{figure*}

In the typical timeline of fit and evaluation \cite{Jacobs2019,Konukoglu2010}, described in 
figure \ref{fig:timeline} as the bidirectional scheme, the model is fitted on a tumor segmentation $S_0$ and then
simulated from onset, through $S_0$, to the point of evaluation $S_2$. In other words, the prediction 
contains the behavior that it is fitted on.

We propose a strictly forward evaluation scheme that separates the model fit from the prediction as much as possible. 
As described in figure \ref{fig:timeline} as the forward scheme, the parameters are fitted on an initial time-point $S_0$ and then used to 
make a prediction between two follow-up scans $S_1$ and $S_2$. By running the prediction from a segmentation instead
of an initial location, the need to fit the initial shape is removed and the evaluation is purely made on 
growth behavior that is unknown when fitting the model. 

For our dataset, we need to consider the role of the tumor resection. In both schemes, the resection cavity as estimated by the aligment of the tissue at $S_0$ and $S_1$, is removed 
from the region of interest for evaluation. In the forward scheme, any voxels in the segmentation $S_1$ are also removed 
from the region of interest, leaving only the new growth visible in $S_2$ for evaluation. So where the bidirectional scheme evaluates
predictive performance on the entirety of the remaining tumor, using $S_0$ only to initialize the location of onset, 
the forward scheme evaluates purely predictive performance based on the knowledge of $S_1$.

\section{Methods}

\subsection{Growth models}
The model under evaluation is a traditional diffusion-proliferation
model with anisotropic diffusion, informed by diffusion tensor imaging (DTI) . 
The cell density $c$ changes with each timestep $dt$ according to:
\begin{equation}
\frac{dc}{dt}=\nabla(\mathbb{D}\nabla c)+\rho c(1-c),
\end{equation}
\begin{equation}
\mathbb{D}\nabla c \cdot n_{\delta\Omega} = 0,
\end{equation}
where $\rho$ is the growth factor, $n_{\delta\Omega}$ is the normal vector
at the brain boundary and $\mathbb{D}$ is a tensor
comprising an isotropic and anisotropic component:
\begin{equation}
\mathbb{D}=\kappa(x)\mathbb{I}+\tau F(x)\mathbb{T}(x),
\end{equation}
where $\kappa$ and $\tau$ are parameters to weigh the two components,
$\mathbb{I}$ is the identity matrix, $F(x)$ is the local Fractional Anisotropy (FA)
and $\mathbb{T}$ is the normalized diffusion tensor \cite{Gholami2016}.

The isotropic diffusion depends on the local tissue type \cite{Jacobs2019}, as
defined by a separate parameter $\kappa_{w}$ and $\kappa_{g}$ for 
voxels in the white matter ($\mathcal{W}$) and grey matter ($\mathcal{G}$) respectively:

\[
\kappa(x)=\begin{cases}
\kappa_{w} & x\in \mathcal{W}\\
\kappa_{g} & x\in \mathcal{G}
\end{cases}
\]

To go from a prediction of $c(t, x)$ to a time-to-invasion ranking $T(x)$, 
a threshold $c_v$ is applied at each iteration such that $T(x) = \min_t c(t,x) > c_v$, where the visibility threshold is set as $c_v = 0.5$. 
The initial condition of the model is provided by an initial cell density $c(t=0)$, 
which is a gaussian distribution centered at a location $x_s$ and a standard deviation of 1mm. 
Alternatively, for the forward evaluation scheme, the model is initialized with a segmentation 
by setting the cell density at $c=c_v$ for voxels inside the segmentation \cite{Elazab2017}.

The model has four unknown parameters ($\rho$, $\tau$, $\kappa_{w}$, $\kappa_{g}$)
and an initial location $x_s$. It was implemented in Fenics \cite{Alnaes2015} in a cubic mesh of 1mm isotropic
cells, using a finite element approach and Crank-Nicolson approximations for the time stepping. 

\subsection{Eikonal approximation}
The behavior of such a growth model can be approximated as a travelling wave, where the speed 
depends only on the local diffusion and proliferation properties. Konokoglu et al. \cite{Konukoglu2010}
have shown that such an eikonal approximation can effectively mimic the evolution of the visible tumor boundary. 
This approximation is especially useful for fitting the initial location $x_s$ of a full growth model \cite{Rekik2013}.
In this work, we use an eikonal approximation that assumes the visible tumor margin moves at a 
speed $v$ of $v = 4 \sqrt{\rho \Tr (\mathbb{D})}$.

The eikonal approximation was used to estimate $x_s$ for 
a given set of model parameters, by optimising the approximation of the initial tumor $S_0$
using Powell's method \cite{Powell2009}. The best approximation was selected from multiple
runs of the optimization with different random seeds.

\subsection{Healthy brain structure}
Running a growth model from onset requires knowledge of the underlying healthy tissue. 
Removing pathology from an image is a research problem 
in itself, but commonly a registration approach with a healthy brain - often an atlas - 
is used \cite{Bakas2016,Lipkova2019,Jacobs2019}. We propose to not use an atlas but rather the 
contralateral side of the brain as a reference for healthy brain structure (similar to \cite{Clatz2005}). 
Using a registration of the T1-weighted image with its left-right mirrored version, all segmentations were transferred to the contralateral
healthy side of the brain. To prevent unrealistic warping of the image due to image intensity changes in the tumor, while still capturing its mass effect, 
the b-spline registration was regularized with a bending energy penalty \cite{Klein2010}.
The weight of this penalty with the mutual information metric was tuned on a number of cases using visual inspection of the transformation. 

The model input is a segmentation of the brain, separated into white matter ($\mathcal{W}$) and gray matter ($\mathcal{G}$), potentially
an estimate of the local diffusion based on Diffusion Tensor Imaging (DTI), and a binary segmentation
of the tumor. Segmentations of the brain and brain tissue were produced using HD-BET \cite{Isensee2019} and
FSL FAST \cite{Zhang2001} respectively. For the pre-operative images, which did not include a T2W-FLAIR sequence, $S_0$ was segmented manually. 
Tumor segmentations $S_1$ and $S_2$ for consecutive images were produced using using HD-GLIO \cite{Kickingereder2019,Isensee2020} 
and corrected manually where necessary. Alignment with the space of $S_0$ was achieved with a b-spline registration,
which was evaluated visually. Datasets were excluded if the registration did not produce a reasonable aligment.

As no registration or segmentation will be perfect, some inconsistencies remain that prevent
a perfect prediction. To not punish the model unfairly, the 
voxels in $S$ falling outside the brain were disregarded in the computation of the Average Precision metric.

\subsection{Patient selection}
A retrospective dataset was selected from Erasmus Medical Center
of patients who a) were diagnosed with a low-grade glioma; b) were treated with surgical resection, but received no chemo- or radiotherapy; and c) had a DTI and 3D T1-weighted 
scan before resection, and two follow-up scans (before and after tumor progression).
This resulted in data of 21 patients, after one dataset was excluded due to failed registration. Note that the time difference between the measurement 
of initial tumor and the two follow-up scans varied from a few months to several years.

\subsection{Parameters}
As the variation of diffusive behavior within the brain is most defining for the tumor shape, we used ${\kappa_{w}}$, ${\kappa_{g}}$ and $\tau$ 
 as parameters of interest, while keeping the proliferation constant at $\rho = 0.01$. These parameters were not fitted
 but rather varied systematically, as listed with the results in \ref{fig:apresults}. For this range of seven growth model parameter settings, the AP performance was 
 measured for goodness-of-fit on the baseline segmentation $S_0$ and predictive performance on $S_2$, according to the two
 evaluation schemes. The relation between goodness-of-fit and predictive performance was quantified using a patient-wise Spearman correlation across different growth model parameter settings.
 The mean of the patient-wise correlation coefficients was tested for a significant difference from zero using a one-sample t-test.

\section{Results}
\begin{figure*}[htb]
    \includegraphics[width=\linewidth]{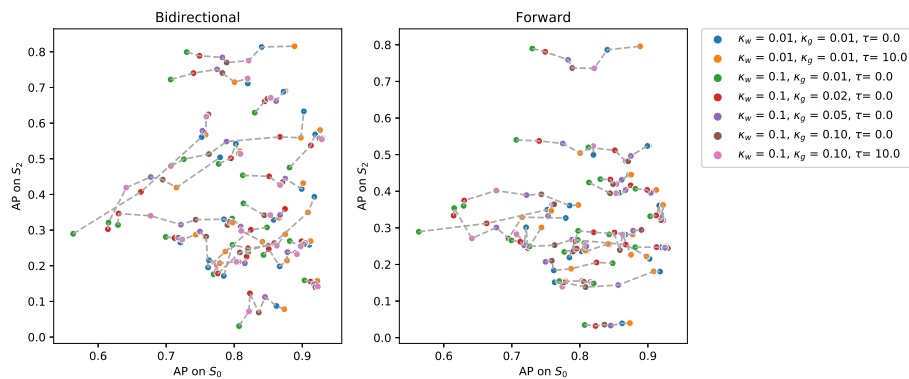}
    \caption{Comparison of goodness-of-fit versus predictive performance for the two evaluation setups. 
    Results for the same patient on different parameter sets are interconnected. }
    \label{fig:apresults}
\end{figure*}

Figure \ref{fig:apresults} shows a comparison of the goodness-of-fit, which is measured by the 
average precision on the initial tumor segmentation $S_0$ and the final predictive performance on $S_2$. 

Comparing the performance between different growth model parameter settings, it is clear that goodness-of-fit is generally
higher and more dependent on the model parameters than the predictive performance. From the growth model parameter settings, typically the best 
goodness-of-fit (AP on $S_0$) was achieved with low diffusion ($\kappa_w = 0.01$) while the worst fit was achieved when the difference in $\kappa$ between
white and gray matter was large ($\kappa_w = 0.1$, $\kappa_g = 0.01$ or $\kappa_g=0.02$).

From the results of the bidirectional evaluation scheme, 
going from an intitial point through $S_0$ to $S_2$, it seems that there is a relation between 
the goodness-of-fit and the predictive performance. However, this relation disappears when using the forward evaluation scheme. 
These observations are confirmed by the mean patient-wise correlation coefficients, which were 0.24 (p=0.06) for the forward scheme and -0.03 (p=0.76)
 for the bidirectional scheme.
\section{Discussion}
This work presents a framing of the tumor growth predictions as a forward ranking problem, and describes the Average Precision metric for 
its evaluation. By formulating the problem in this way we can evaluate the full spatiotemporal results of the 
results, even if the observations are only snapshots in the form of a segmentation. A further advantage is found in the 
direct link to local growth speed and quantification of the local model agreement. Though these advantages are only of a qualitative nature,
and do not provide a direct benefit to the model itself, we believe it to be a useful step towards the improvement and better comparison
of tumor growth predictions. An important underlying assumption in this framework is that the time axis is not 
quantified, so the prediction does not provide information on the overall speed of growth or any potential mass effect. Predicting 
the growth speed is a highly relevant problem as well, but very different in nature from the spatial distribution. 

The importance of framing the problem is further illustrated with the experimental setup. Specifically for personalized 
tumor growth models, which are fitted to an initial tumor shape, this work presents 
a forward scheme that separates the goodness-of-fit from the evaluation of future predictions. Using this scheme, and 
comparing to a bidirectional alternative, we show that the goodness-of-fit and predictive performance are not necessarily correlated. 
However, we should be careful in interpreting these results as the underlying reason for the disparity 
is unclear. If the goal is to find the model that best fits the available data, this strict separation might not be useful. 

When moving towards statistical models for tumor growth, the framing of the problem is essential. Between the actual
mechanisms of tumor growth and the segmentation is a flawed observation on MR imaging, the rather difficult problem of segmentation 
and registration and an estimate of the time horizon. Those factors, combined with limited data and the fact that glioma are naturally
unpredictable are a major reason why tumor growth models have relied heavily on simulations \cite{Ezhov2019} and qualitative observations \cite{Angeli2018} for their validation. 
This work is a step towards the comparison and clinical evaluation of tumor growth predictions that fits their spatiotemporal nature,
and allows for localized interpretation.

\section*{Acknowledgements}
This work was supported by the Dutch Cancer Society (project number 11026,
GLASS-NL) and the Dutch Organization for Scientific Research (NWO).
\FloatBarrier

%
%
%
\bibliographystyle{splncs04}
\bibliography{miccai_library}
\end{document}